\documentclass{article}
\usepackage{spconf,amsmath,graphicx,amssymb}
\usepackage{enumitem}
\setlist{nosep, leftmargin=14pt}
\usepackage{comment}
\usepackage{bm}
\usepackage{mwe} 
\usepackage{xcolor}
\usepackage{flushend}

\graphicspath{{images/}}

\def\x{{\bm{x}}}
\def\y{{\bm{y}}}
\def\n{{\bm n}}
\def\z{{\bm z}}
\def\H{{ \bm{\mathcal H}}}
\def\F{{\bm{\mathcal F}}}

\title{Display Field-of-View Agnostic Robust CT Kernel Synthesis \\using Model-Based Deep Learning}

\name{Hemant Kumar Aggarwal$^{\star}$, Antony Jerald$^{\star}$, Phaneendra K.~Yalavarthy$^{\star, \dagger}$, Rajesh Langoju$^{\star}$, and Bipul Das$^{\star}$}

\address{$^{\star}$Science and Technology Organization, GE HealthCare, Bangalore, India \\
		$^{\dagger}$Department of Computational and Data Sciences, Indian Institute of Science, Bangalore, India}

\begin{document}

\maketitle

\begin{abstract}
	
In X-ray computed tomography (CT) imaging, the choice of reconstruction kernel is crucial as it significantly impacts the quality of clinical images. Different kernels influence spatial resolution, image noise, and contrast in various ways. Clinical applications involving lung imaging often require images reconstructed with both soft and sharp kernels. The reconstruction of images with different kernels require raw sinogram data and storing images for all kernels increases processing time and storage requirements. The Display Field-of-View (DFOV) adds complexity to kernel synthesis, as data acquired at different DFOVs exhibit varying levels of sharpness and details. This work introduces an efficient, DFOV-agnostic solution for  image-based  kernel synthesis using model-based deep learning. The proposed method explicitly integrates CT kernel and DFOV characteristics into the forward model. Experimental results on clinical data, along with quantitative analysis of the estimated modulation transfer function using wire phantom data, clearly demonstrate the utility of the proposed method in real time. Additionally, a comparative study with a direct learning network, that lacks forward model information, shows that the proposed method is more robust to DFOV variations.
\end{abstract}
\begin{keywords}
X-ray Computed Tomography, Kernel Synthesis, Model-based Deep Learning
\end{keywords}
\section{Introduction}
\label{sec:intro}
X-ray computed tomographic (CT) image Kernel Synthesis (KS) involves transforming an image reconstructed with one kernel into an image reconstructed with another kernel~\cite{radiologyct,imageQuality,ohkubo2016Filtering}. This process has various applications, including improving low contrast distinguishability, enhancing computer-aided detection~\cite{icnodulenet}, and facilitating quantitative analysis~\cite{tanabe2022kernel}.

\begin{figure}
	\centering
	\includegraphics[width=.6\linewidth]{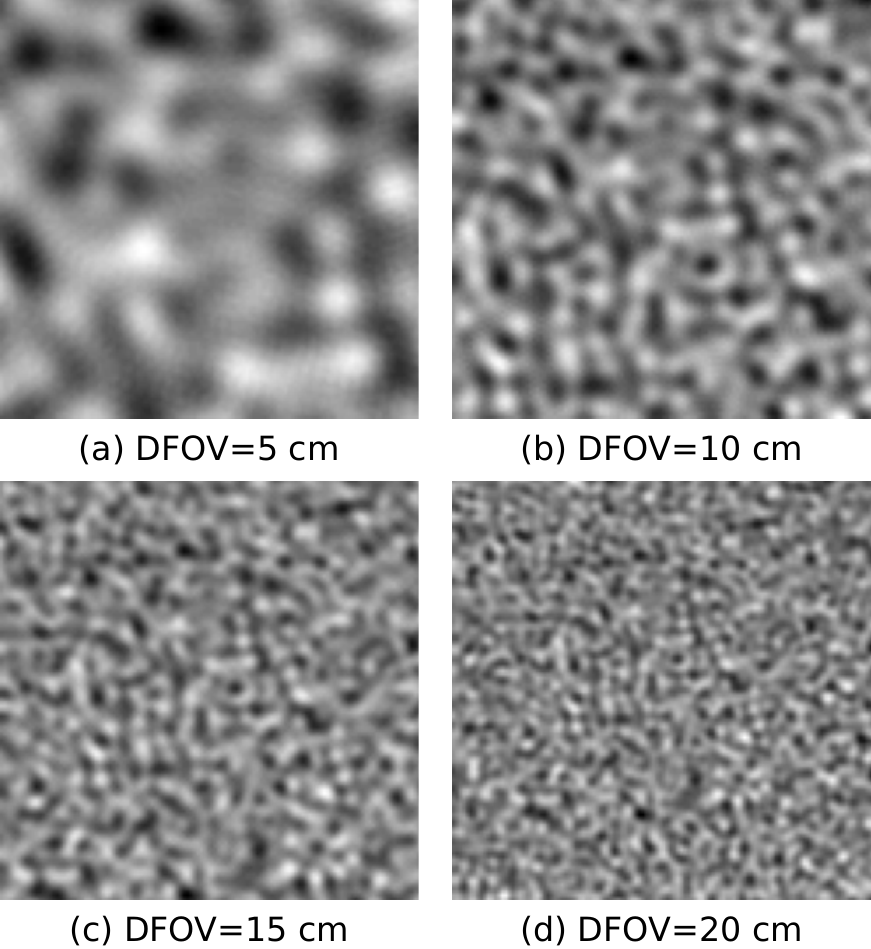}	
	\caption{ A $100 \times 100$ patch of a water phantom scanned at different DFOV values demonstrates significant texture changes as the DFOV increases. This dependency makes it challenging to develop a image-based kernel synthesis model that is agnostic to DFOV. }
	\label{fig:wp}
\end{figure}

When reconstructing an image from raw sinogram data using the filtered back projection (FBP) algorithm, a reconstruction kernel is typically used to emphasize certain anatomical regions depending on the application. For instance, smooth kernels offer excellent low-contrast distinguishability with lower noise and fewer artifacts but have low spatial resolution. Conversely, sharp kernels provide high spatial resolution but also increase noise and artifacts. If raw sinogram data is unavailable or not stored due to large storage requirements, image-based kernel synthesis is used to reconstruct images with different kernels.
\begin{figure}
	\centering
	\includegraphics[width=.9\linewidth]{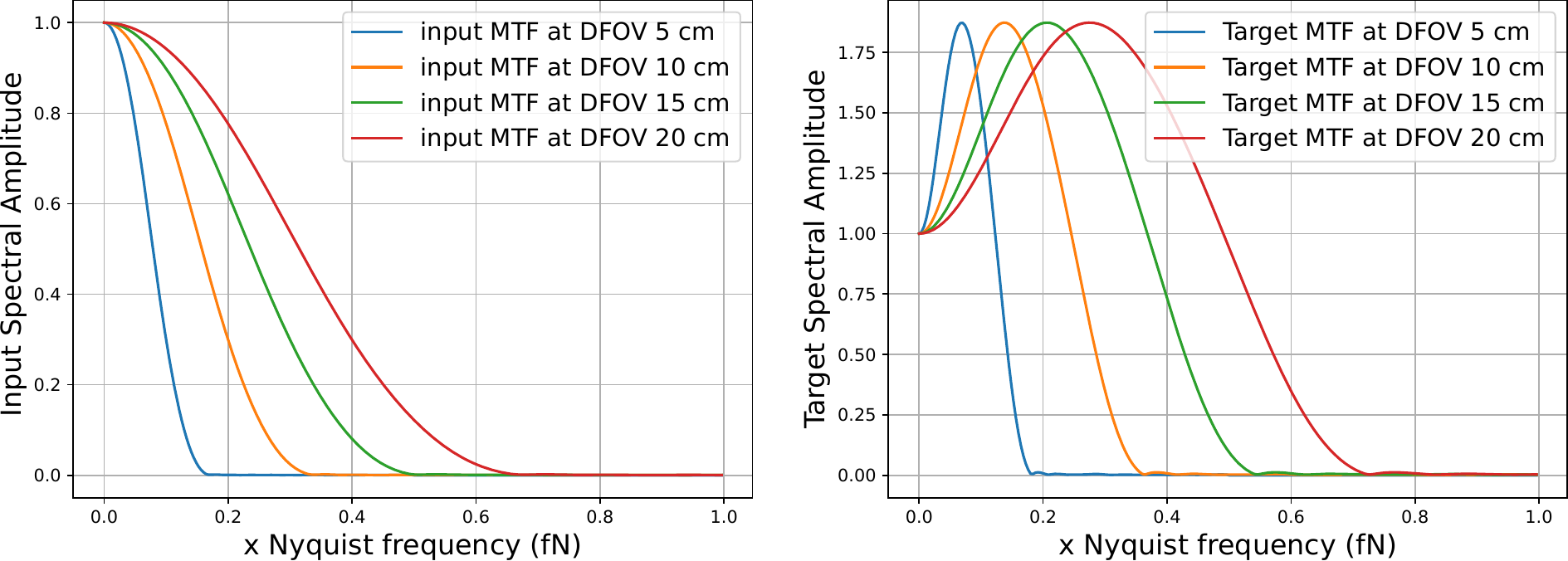}	
	\caption{Input (smooth) and target (sharp) kernel MTFs corresponding to different DFOVs. An increase in DFOV leads to high frequency content in the image. }
	\label{fig:mtfs}
\end{figure}
However, image-based kernel synthesis remains a challenging inverse problem, as the synthesis must be robust to variations in the display field of view (DFOV) of the reconstructed images. For instance, Figure~\ref{fig:mtfs} illustrates the modulation transfer functions (MTFs) of smooth (input) and sharp (target) kernels at different DFOVs, ranging from 5 cm to 20 cm. It is evident from the plots that varying DFOVs result in different resolutions of reconstructed images. The effect of DFOV on image quality (IQ) is further demonstrated in Figure~\ref{fig:wp} using real water phantom data, where an increase in DFOV corresponds to increased speckle sharpness. In addtion to DFOV variations, managing artifacts and noise is another challenge in image-based kernel synthesis, especially when converting from smooth to sharp kernel images.
Ohkubo et al.~in~\cite{ohkubo2011image} proposed a direct kernel synthesis method that employs the ratio of modulation transfer functions (MTFs) between input and target kernels to transform images. This method is inherently robust to variations in the display field of view (DFOV) but often results in decreased image quality (IQ), particularly when converting smooth to sharp kernel images, due to amplified noise and artifacts in the reconstructed images.

Recently, several deep learning methods have been proposed to address kernel synthesis problem. Unlike direct kernel synthesis methods, these approaches are effective at managing noise during image transformation but struggle with variations in DFOVs. For example, a simple direct learning method like U-Net~\cite{unet}, trained on a specific DFOV, does not perform well when applied to data with different DFOVs. Similarly, experiments show that a network trained on data from all DFOVs tends to bias towards a specific DFOV.

To overcome this issue, we propose a model-based deep learning method that explicitly incorporates kernel MTFs into the deep learning framework. The proposed architecture for solving kernel synthesis problem consists of two parts: a deep learning-based projection step and an iterative data consistency solver that utilizes these MTFs. Enforcing data consistency during the training of the denoising network makes the kernel synthesis solution DFOV-agnostic, allowing a single network to be trained across various DFOVs with the advantage of requiring only a few hundred training samples.

This work complements our previous research~\cite{aggarwal2024self} by developing a deep model in a supervised manner that is agnostic to DFOV  using clinical data whereas previous work~\cite{aggarwal2024self} focused on self-supervision using single-slice for noise aware kernel synthesis. A continuous kernel synthesis approach has been discussed in~\cite{jongContinuousConversion} but it is self-supervised. In contrast to~\cite{jongContinuousConversion}, our proposed method is supervised learning with focused on DFOV and CT noise awareness.

\section{Proposed Method}
\begin{figure}
	\centering
	\includegraphics[width=1\linewidth]{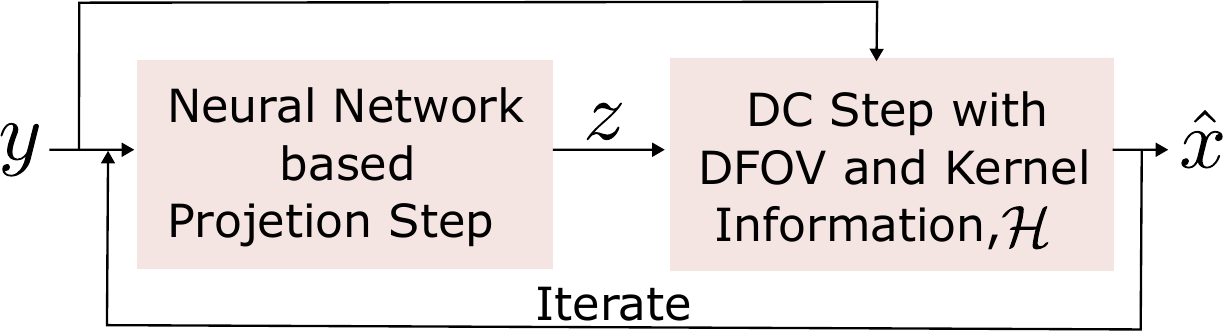}	
	\caption{The proposed training pipeline explicitly utilizes DFOV dependent water phantom data as noise together with input and target slice pairs. The network based project step acts as a generic DFOV agnostic denoiser that is shared across five unrolls used during network training. The analytical solution to the Data Consistency (DC) step is shown in Eq.~\eqref{eq:dc}. Incorporating the DC step explicitly into the learning framework helps in developing  DFOV agnostic deep model.}
	\label{fig:proposed_method}
\end{figure}

The image-domain kernel synthesis can be expressed as a linear inverse problem of the form
\begin{equation}
	\label{eq:forward}
	\y =  \H  \x+ \n,
\end{equation}
where $\y \in \mathbb{R}^N, N=p \times q$ represents vectorized input kernel image of $p$ rows and $q$ columns, $\x \in \mathbb{R}^N$ is the target kernel image which needs to be synthesized, $\H$ is the forward operator representing kernel synthesis process having input and target kernel information at specific DFOV. More specifically, using diagonalization, $\H$ can be expressed as  $\H=\F^T \bm{\Lambda} \F$ where, $\F$ is the Fourier transform operator, $ \bm{\Lambda}$ represents a DFOV dependent ratio of input and target CT kernel's modulation transfer function~(MTF). Unlike tradition inverse problems in imaging, $\n$ is a kernel and DFOV dependent noise that is assumed to be additive, as discussed in Fig.~\ref{fig:wp}.  

\begin{figure*}
	\centering
	\includegraphics[width=.9\linewidth]{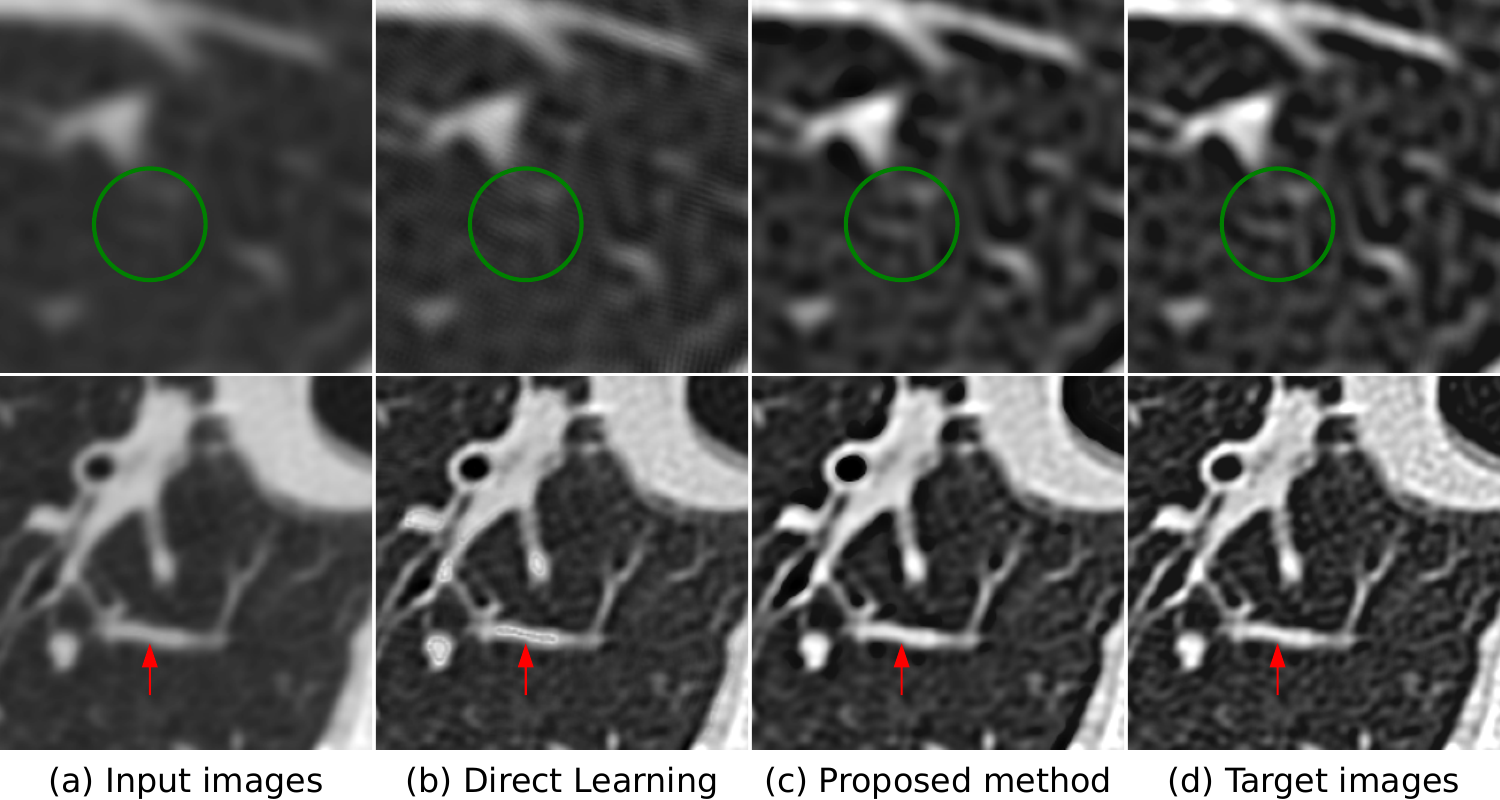}	
	\caption{The top row shows inference results at a DFOV of 5 cm, and the bottom row at a DFOV of 10 cm. The proposed method in~(c) produces sharper output compared to the direct learning  method in (b). The green circle highlights artifacts (in zoomed version) in the DFOV 5 cm output of the direct learning method. The red arrow indicates hallucinations in the direct learning method, whereas the proposed method retains the structure well. }
	\label{fig:results}
\end{figure*}

It is possible to utilize a direct learning approach such as UNet that takes $\y$ as input and $\x$ as target and learns a mapping $\hat{\x}=\text{CNN}(\y)$ between input and target pair of images to result in network prediction $\hat{\x}$. However, such a direct learning approach depends on the large amount of training dataset consisting of DFOV variation which is difficult to collect for x-ray CT imaging. Further, without the explicit information about the Kernel's MTF and DFOV, the learned model does not necessarily leads to data-consistent result as demonstrated in the experimental results in Fig.~\ref{fig:results}. 

This work propose to find a solution to kernel synthesis inverse problem~\eqref{eq:forward} using model-based deep learning~\cite{modl} and represents the network regularized optimization problem in the form
\begin{equation}
	\label{eq:modl}
	\arg\min_{\x \z} ||\y - \H \x ||_2^2 + \lambda ||\x -\z||^2_2,
\end{equation}
where, $\z$ represents the output of a CNN that satisfy data consistency constraint as explicitly implied by the above problem formulation. Here, $\lambda $ is a regularization parameter initialized with $0.5$ and decayed with iteration, $\lambda_{k+1} =0.9\lambda_k$, to gradually give more importance to data consistency term. The above problem~\eqref{eq:modl} can be solved iteratively in two steps using alternating minimization as follows:
\begin{align}	
	\z_k&= \text{CNN}(\x_k) \label{eq:cnn}\\
	\x_{k+1}&= \F^T \left( \frac{\F(\H^T \y + \lambda \z_k)	}{|\bm{\Lambda} |^2 + \lambda} \right ) \label{eq:dc}, 	
\end{align}
where $\x_0$ can be initialized as Tikhonov solution of the form $\x_0=(\H^T \H +\lambda \bm{\mathcal I} )^{-1} \H^T \y$, here $\bm{\mathcal I}$ is identity matrix. The analytical solution in Eq.~\eqref{eq:dc}  assumes $\z_k$ to be constant for the data-consistency problem.

As visually demonstrated in Fig.~\ref{fig:wp}, DFOV impacts the texture of water phantom data, which is used as noise during network training. Fig.~\ref{fig:proposed_method} presents a schematic diagram of the end-to-end network training strategy using supervised learning with existing clinical data from different DFOVs. The use of DFOV-dependent water phantom data as noise allows for controlling the enhancement of noise and artifacts during the iterative reconstruction process.

\section{Trainings and Experiments}

Two separate supervised training setups were used. The first was a direct learning method that trained a U-Net to perform combined DFOV training without kernel MTF information. The second was the proposed method, which employed a U-Net with weight sharing across five unrolls and incorporated explicit DFOV and kernel information in the forward model. 

The networks were trained using existing clinical lung images obtained from the GE Revolution Ascend system, acquired with 120 kVp and 544 mA, and reconstructed with different DFOV values of 5 cm, 10 cm, 15 cm, and 20 cm. The training data consisted of 1280 slices, each of size $512 \times 512$, from different subjects. STANDARD kernel images were used as input, and LUNG kernel images were used as target images for all DFOVs. Both networks were trained for 500 epochs with an initial learning rate of $1 \times 10^{-4}$, using mean square error (MSE) combined with the structural similarity index (SSIM) as the loss function. Following the training, the performance of the trained networks was verified using both clinical and phantom data across various DFOV.

In the first experiment, we compared the reconstruction quality of the proposed method with that of the direct learning method. Fig.~\ref{fig:results} shows portions of the input and target kernel images, along with the outputs of the direct learning method and the proposed method for DFOV 5 cm and DFOV 10 cm, respectively, in the top and bottom rows. The proposed method achieves sharpness closer to the ground truth while preserving structural information, unlike the direct learning method, which results in blurred images with hallucinations, as indicated by the green circle and red arrow.

\begin{figure}
	\begin{minipage}[b]{0.35\linewidth}
		\centering
		\centerline{\includegraphics[width=1\linewidth]{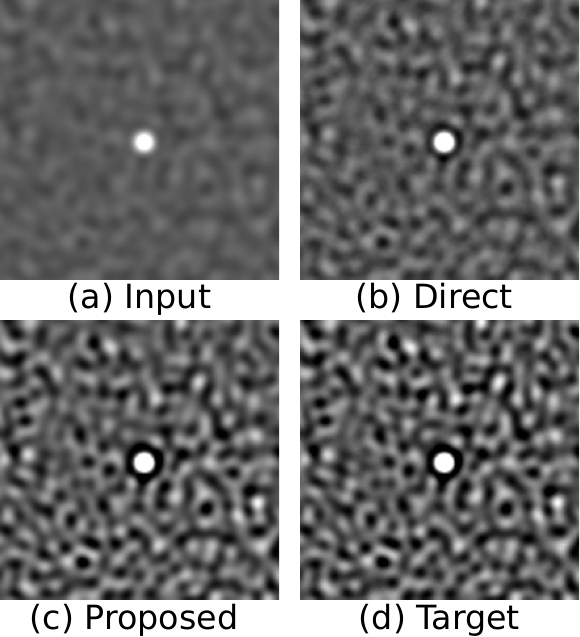}	}		
		\centerline{(a) Wire phantom }\medskip
	\end{minipage}	
	\begin{minipage}[b]{0.62\linewidth}
		\centering
		\centerline{\includegraphics[width=1\linewidth]{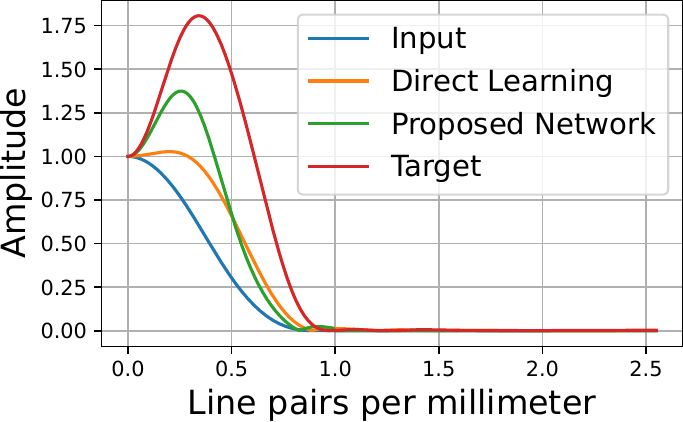}}		
		\centerline{(b) Estimated MTFs at DFOV 10~cm }\medskip
	\end{minipage}
	\hfill
	\caption{(a) Estimated wire phantom images visually show that proposed method results in sharper output compared to a direct learning method. (b) Quantitative results at DFOV 10 cm on wire phantom data demonstrate that the proposed method produces sharper output compared to the direct learning method. The orange curve represents the direct learning method, which has less mid-frequency boost compared to the green curve representing the proposed method.}
	\label{fig:mtfestimation}
\end{figure}

In the second experiment, we quantitatively evaluated the performance of the proposed method by estimating the MTF on the DFOV 10 cm wire phantom data. Figure~\ref{fig:mtfestimation}(a) shows the wire phantom images for the input, target, direct learning method, and proposed method. It is evident from the images that the wire image produced by the proposed method is closer to the target wire image compared to the wire image from the direct learning method. Figure~\ref{fig:mtfestimation}(b) displays the MTFs estimated for the proposed method and the direct learning method, compared with the input and target MTFs. The estimated MTF by the proposed method shows a higher mid-frequency boost compared to the direct learning method and is closer to the target.

\section{Conclusions}

This work presented a DFOV-agnostic image-based kernel synthesis method that performs kernel synthesis which is robust to input noise. This method significantly improves disease diagnosis and treatment by enhancing the visibility and precision of anatomical structures such as bones, vessels, and lesions as shown by converting smooth kernel images to sharp kernel images. 

\section{Compliance with ethical standards}
\label{sec:ethics}
This research study was conducted using human subject data made available through appropriate research contracts. All the activities necessary for obtaining the research results/development were carried out with due approval from the ethics committee and competent authorities in compliance with the regional laws and regulations.

\bibliographystyle{IEEEbib}

\end{document}